\documentclass[sigconf]{acmart}
\AtBeginDocument{%
  \providecommand\BibTeX{{%
    \normalfont B\kern-0.5em{\scshape i\kern-0.25em b}\kern-0.8em\TeX}}}

\copyrightyear{2023}
\acmYear{2023}
\setcopyright{rightsretained}
\acmConference[CHI EA '23]{Extended Abstracts of the 2023 CHI Conference on Human Factors in Computing Systems}{April 23--28, 2023}{Hamburg, Germany}
\acmBooktitle{Extended Abstracts of the 2023 CHI Conference on Human Factors in Computing Systems (CHI EA '23), April 23--28, 2023, Hamburg, Germany}\acmDOI{10.1145/3544549.3582735}
\acmISBN{978-1-4503-9422-2/23/04}

\usepackage{tfrupee}
\usepackage{epigraph}

\begin{document}

\title{Enough With ``Human-AI Collaboration''}

\author{Advait Sarkar}
\affiliation{%
  \institution{University of Cambridge and University College London}
  \country{United Kingdom}
}


\begin{abstract}
  Describing our interaction with Artificial Intelligence (AI) systems as `collaboration' is well-intentioned, but flawed. Not only is it misleading, but it also takes away the credit of AI `labour' from the humans behind it, and erases and obscures an often exploitative arrangement between AI producers and consumers. The AI `collaboration' metaphor is merely the latest episode in a long history of labour appropriation and credit reassignment that disenfranchises labourers in the Global South. I propose that viewing AI as a tool or an instrument, rather than a collaborator, is more accurate, and ultimately fairer.
\end{abstract}

\ccsdesc[500]{Human-centered computing~HCI theory, concepts and models}
\ccsdesc[300]{Computing methodologies~Artificial intelligence}
\ccsdesc[300]{Computing methodologies~Machine learning}
\ccsdesc[300]{Social and professional topics~Intellectual property}
\ccsdesc[300]{Social and professional topics~Surveillance}
\ccsdesc[300]{Social and professional topics~Geographic characteristics}
\ccsdesc[300]{Social and professional topics~Race and ethnicity}

\keywords{metaphors, labour, fairness, equity, datasets, labelling, annotation}

\begin{teaserfigure}
  \includegraphics[width=0.5\textwidth]{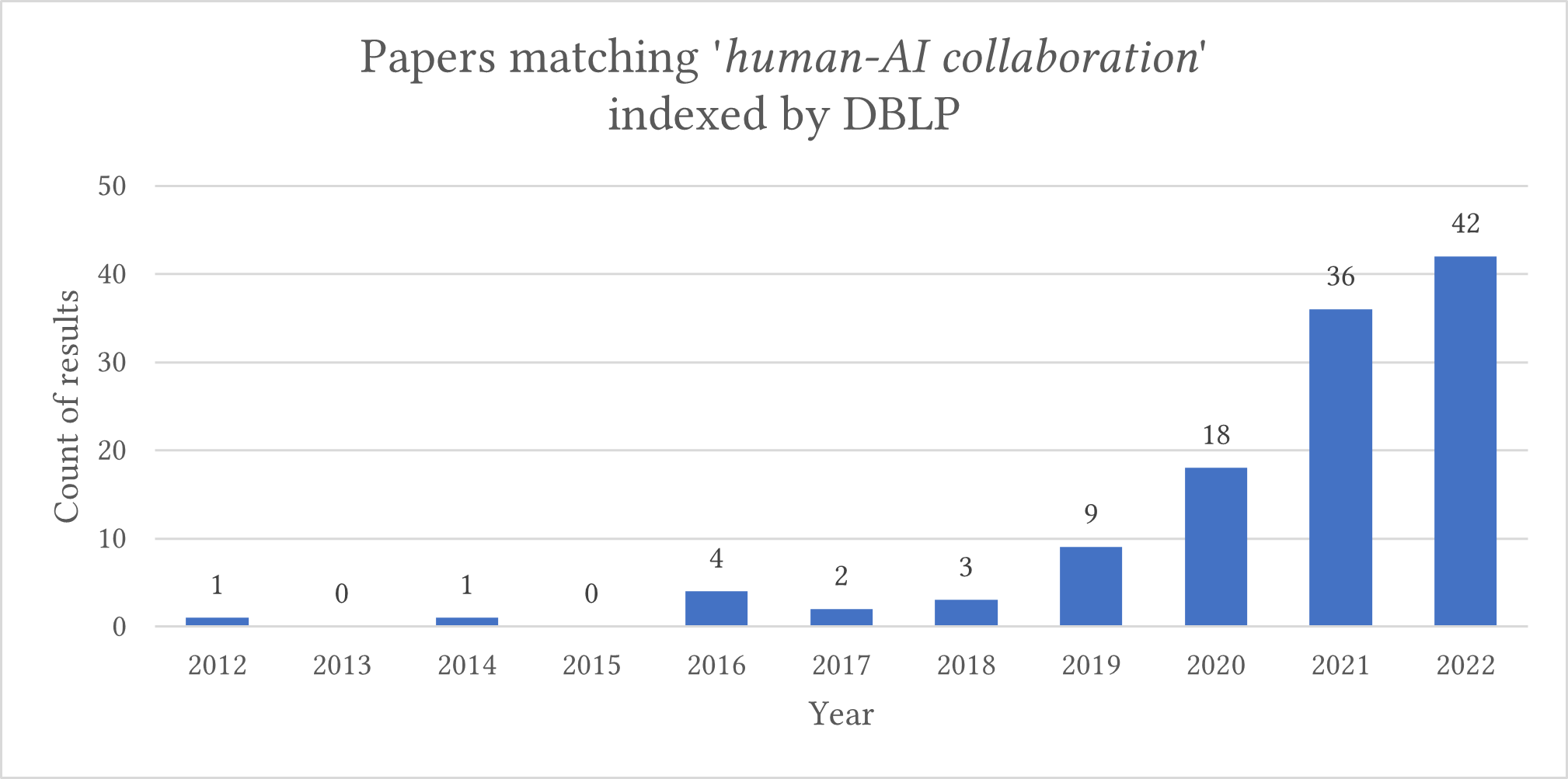}
  \includegraphics[width=0.5\textwidth]{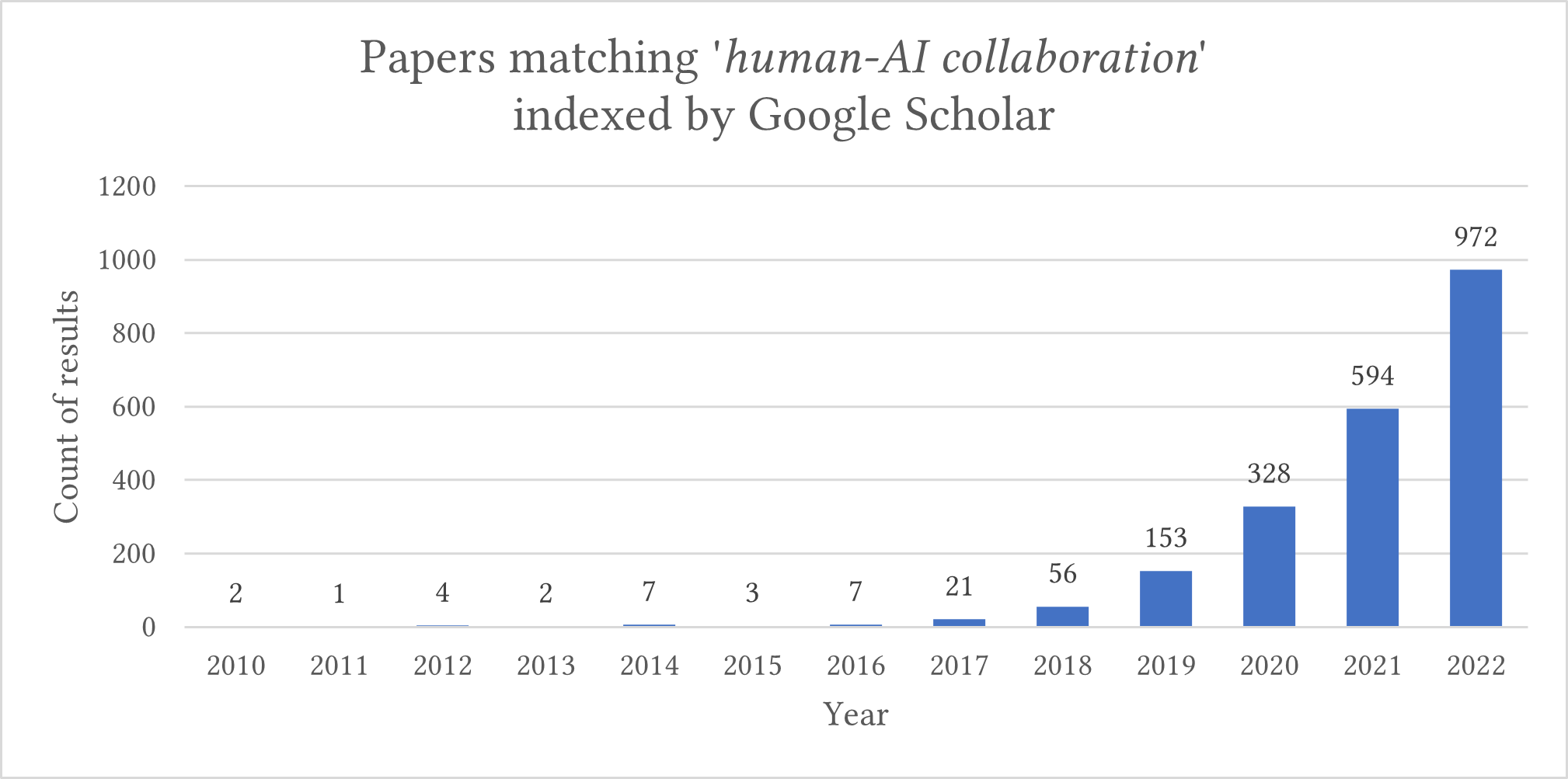}
  \caption{The use of the phrase `human-AI collaboration' in academic scholarship has dramatically increased in the last 5 years. Left: Results for `human-AI collaboration' on DBLP; Right: results for `human-AI collaboration' on Google Scholar. Figures as of 11 December 2022.}
  \Description{Two time series bar charts, each with several bars representing the count of results in a certain year. On the x-axis is years from 2015 to 2022. On both charts the count increases dramatically as the years go on.}
  \label{fig:citations}
\end{teaserfigure}


\maketitle

\section{The agentistic turn of human-computer interaction}



In May 2022 I attended CHI in New Orleans. As ever, there was a healthy interest in Artificial Intelligence; Grudin's prediction of the (re)convergence of AI and HCI \cite{Grudin_2009} has largely come to pass. An increasingly common perspective in the papers, presentations, and conversations around CHI was that interaction with intelligent systems is deeper, more profound than interaction with `simpler' systems. CHI delegates heralded AI systems as ``teammates'', spoke of our ``partnerships'' with AI, and of ``human-AI collaboration''. This conferral of personhood and agency to computer programs can be thought of as an \emph{agentistic turn} in HCI research. 


The agentistic turn is a widespread phenomenon that goes beyond any individual author or group. For example, papers matching the query ``human-AI collaboration'' have been steadily increasing on the computer science bibliography website DBLP. As of 11 December 2022, Google Scholar has indexed over 2,250 papers containing the exact phrase ``human-AI collaboration'' (Figure~\ref{fig:citations}).   



As I walked through the streets of New Orleans every morning to the conference center, the deep history of Louisiana seemed to radiate from its neighbourhoods, its architecture, its food, and its people. Shaped by French, Spanish and American colonial powers over 300 years, Louisiana's history and prosperity is built, perhaps more than any other American state's, on slavery.

While in New Orleans, I visited Oak Alley, a sugar plantation from the mid-1800s known for its picturesque oak boulevard connecting the plantation's stately manor to the former banks of the Mississippi (the river has since been moved to make room, in characteristically American fashion, for a highway). Oak Alley, like many plantations in the region, exploited the work of hundreds of slaves in the dangerous and backbreaking work of sugar cultivation and production.

\begin{figure}[t]
  \centering
  \includegraphics[width=\linewidth]{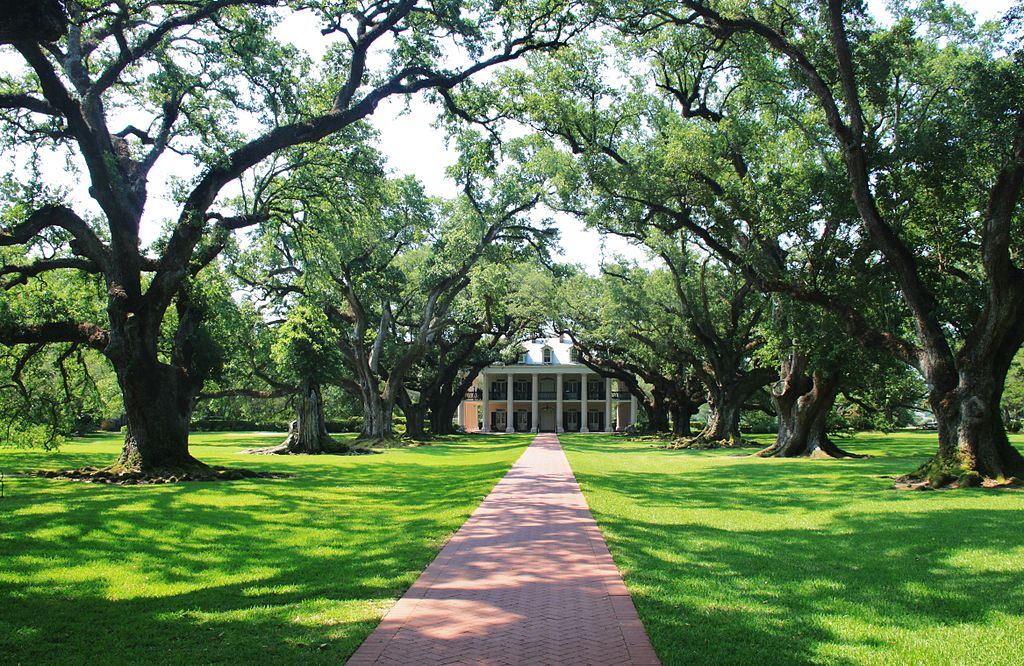}
  \caption{
    Oak Alley plantation, Louisiana, where the slave gardener Antoine developed the papershell pecan.\\Source: \href{https://creativecommons.org/licenses/by-sa/3.0}{RonPaul86, CC BY-SA 3.0}, via \href{https://commons.wikimedia.org/wiki/File:Oak_Alley_Plantation.JPG}{Wikimedia Commons}
  }
  \label{fig:oak_alley}
  \Description{A red brick footpath leads to a stately manor. On either side is green grass and a row of mature oak trees.}
\end{figure}


I was taken by the story of one such slave, known to us only as Antoine. A remarkably talented gardener, Antoine discovered a process of grafting pecan trees, which grew wild in the region and had been harvested and enjoyed as food since the pre-colonial Native Americans, but which had previously been impossible to cultivate. Through his efforts, he developed the `papershell' pecan, whose fragile shell could be cracked by hand, and this property, together with his newly perfected grafting technique, unlocked industrial-scale pecan production and consumption \cite{wells_2017, shannon_2021}. A later owner of the plantation submitted these pecans to the 1876 Centennial Exhibition in Philadelphia, where they won an award and their formal moniker the `centennial'.

Pecans and sugar (and their combined confection, the praline) are icons of Louisiana; New Orleans' touristy French Quarter overflows with the stuff (I remarked to my fellow conference delegates that one can scarcely open one's mouth in New Orleans without a beignet or a praline spontaneously flying into it). Thus, we enjoy the labour of this nearly anonymous slave, who himself enjoyed neither prosperity nor recognition for his outstanding ingenuity and contribution.

\section{The hidden work of artificial intelligence}

\epigraph{Who built Thebes of the seven gates?\\In the books you will find the names of kings.\\Did the kings haul up the lumps of rock?}{Bertolt Brecht, \emph{Questions From A Worker Who Reads}}

\subsection{The data annotation industry: AI's ``back office''}
Knowledge work in the 21\textsuperscript{st} century is so hypersaturated with the word `collaboration', it can be easy to forget that the root of the word is `labour'. To collaborate, is to co-labour, is to labour together. But in a `human-AI collaboration', whose labour are we talking about? Is the portion of labour attributed to AI purely digital in origin, and do computers directly transform energy into work?

The labour sharing and credit assignment of most software is fairly transparent: the programmer labours to write code, and is compensated and credited for this act. The software user labours to use this software, and is compensated and credited for what they produce with it. Programmers at Adobe build Photoshop, they get credit for building the software, and they get paid for it. Designers and artists use Photoshop to do their work, and they get payment and credit for doing this work. The same is true of spreadsheets, word processors, databases, and so on.

This is the `tool' model of credit assignment. The relationship between programmer, software and user, mirrors the relationship between the camera manufacturer, the camera, and the photographer. Or the blacksmith, the chisel, and the sculptor. This relationship involves the creative human labour of two principal parties: the toolmaker and the tool user.

Artificially intelligent systems, however, include a third party: the data labeller. With few exceptions, almost every AI system relies on labelled training data, that is, examples of how to do the job it is supposed to do. Need a face recognition AI? You need a dataset of images with the faces labelled. Need an AI that can sort out defective products on an assembly line? You need a dataset where humans have indicated examples of what a defective product looks like, and what a normal product looks like. Need an AI that can flag harmful content on social media? You need a dataset where harmful content has been identified as such by humans, and so on. Rather than have rules for intelligent behaviour directly coded into these systems by programmers, these systems infer rules for behaviour based on the demonstrations of human behaviour embodied in these training datasets.

To respond to the growing need for labelled training data, a billion-dollar (and rapidly growing) global industry has emerged. A 2021 Gartner report identifies labelling as a key factor for AI development \cite{wesley_2021}. Leading solutions include Scale AI and Amazon Mechanical Turk.

Where do data labelling workers come from? Unsurprisingly, this type of work skews heavily towards the poor countries of the Global South. India is known as AI's ``back office'', where entrepreneurs set up shop in tiny villages, drawing on the inexpensive labour pools~\cite{murali_2019}. When Venezuela's economy collapsed, thousands of desperate Venezuelans entered the industry seeking to earn prized US currency~\cite{jones_2021}. Vulnerable workers around the world, from Kenya to the Philippines, power the artificial intelligence industries of the West, for less than USD \$30 a week~\cite{elliott_2021}.

\begin{figure}[t]
  \centering
  \includegraphics[width=\linewidth]{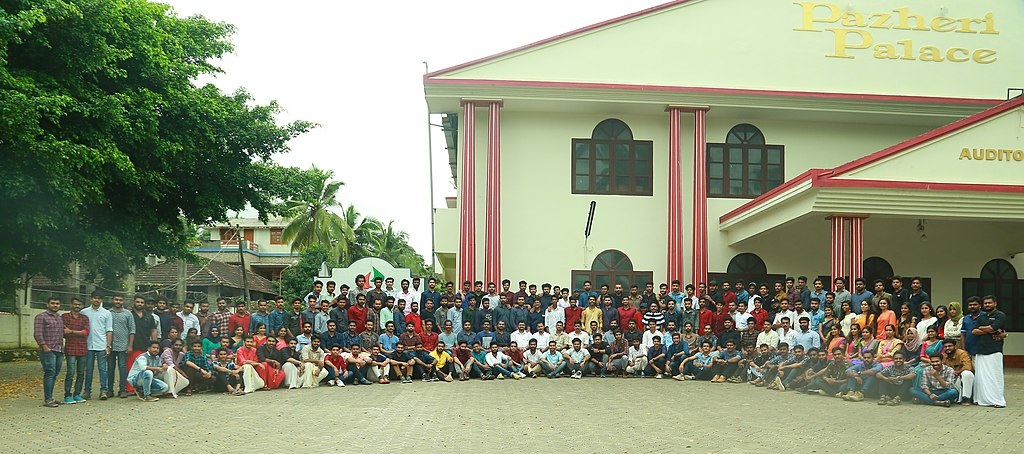}
  \caption{
    Employees of the Indian data labelling company Infolks, based in Kumaramputhur, Kerala. The starting monthly wage is INR \rupee15,000 (around USD \$200 at the time of writing). Source: \href{https://creativecommons.org/licenses/by-sa/4.0}{INFOLKSMKD, CC BY-SA 4.0}, via \href{https://commons.wikimedia.org/wiki/File:Infolks_employees.jpg}{Wikimedia Commons}
  }
  \label{fig:infolks}
  \Description{About one hundred employees sit or stand in four rows for a group photograph on a courtyard in front of a neoclassical office building. Only the left half of the building is visible. Trees are visible in the background. The text "Pazheri Palace" and "Audito" are visible on the building.}
\end{figure}


The daily work of data annotation is gruelling. For example, TikTok moderators are asked to review as many as 1,000 videos a day, causing them to watch videos at 300\% the normal speed~\cite{deck_2022}. Often, these videos are in languages they do not know. Investigation by \emph{Time} magazine discovered that the virally successful ChatGPT application relied on toxicity filters trained by Kenyan workers paid less than USD \$2 per hour, a harrowing experience that left many annotators with lasting mental health issues and described by one annotator as \emph{``torture''} \cite{perrigo_2023}.

AI developers have been interviewed about their relationships with the field workers who provided the labelled data for their systems~\cite{10.1145/3491102.3517578}. Labelling data often requires deep domain expertise; field workers included health workers (radiologists, oncologists, etc.), farmers, wildlife patrollers, oceanologists, teachers, and drivers. These domain experts are called upon to annotate data for AI systems that aim to automate part of the work they do.

Rather than knowledge partners, the AI developers viewed these workers as no more than mere datasets, and judged them as ``corrupt'', ``lazy'', and ``non-compliant''. Rather than discussing and engaging with their expertise, the developers looked to surveillance, gamification, incentives, and punitive measures to improve the data quality ``output'' from these field workers. Ultimately, this reduced the scope for the sophisticated and nuanced expression of expert judgements. If the British East India Company forced Bengali silk artisans to cut off their own thumbs~\cite{bolts1772considerations} thus deskilling them, then is the industrialised deskilling of knowledge work akin to the lobotomization of knowledge workers?

The founder of Indian labelling company Infolks would prefer not to view labelling as knowledge work. In an interview~\cite{mehrotra_2022} he remarks: \emph{``Take the Burj Khalifa in Dubai. The engineer built the plan. But it was thousands of labourers over many years that made it possible [...] The whole credit goes to the engineer who designed it, but without the labourers, he cannot build. We don't care. For working people, they are getting paid. They are satisfied. If they are fine, then why do we need to think about it?''} His error, perhaps deliberate, is that construction work is a poor analogy for annotation work.\footnote{And indeed, we ought to further question whether the construction labour credit sharing status quo is fair or acceptable.}

Labelled training data is not an undifferentiated, commodified input into the software development process (unlike the electricity used to power the programmer's computers, or the food programmers eat to power their bodies). It is bespoke, requires intentional creative acts from data labellers, and is highly directed towards the specific goals of the system being built. 

Reductionist approaches to data annotation are not only a problem for the lives of data workers, they can even poison the results of models and therefore threaten the validity of the entire enterprise. The policies of platforms such as Amazon Mechanical Turk often allow employers to reject the work of an annotator and refuse payment with little room for appeal or recourse. With such a threat, it is no surprise that annotators are strongly influenced to follow the instructions provided, which may have been written without due consideration or understanding of the nuances of the annotation task, and are a source of bias~\cite{https://doi.org/10.48550/arxiv.2205.00415}. Annotators can bring their own biases, such as racial biases. For example, AI systems trained on datasets annotated for hate speech and abusive language showed systematic racial bias, and were substantially more likely to classify tweets written in African-American Vernacular English as abusive~\cite{davidson-etal-2019-racial}.

Gray and Suri use the evocative term ``ghost work'' to describe the invisible labour embodied in these technologies~\cite{gray2019ghost}. Technology firms have two incentives to keep such work invisible: it allows them to cultivate an aura of technological `magic' for the consumer, and maintains the illusion of scalability for investors.

The fact that the data labelling industry relies on vulnerable workers from the Global South is often presented as a positive phenomenon, an opportunity for ``social inclusion and mobility''~\cite{vargas_2019}, and for workers to get experience and a foothold in the lucrative technology sector, a stepping stone to greener pastures. Alas, this rosy picture is false.

In reality, these roles offer very little mobility. Even as some were presenting CHI 2022 papers on human-AI ``collaboration'', down the hall Wang et al. were presenting their research on the aspirations of data annotators in India~\cite{10.1145/3491102.3502121}. There, the industry has become a veritable arms race, to such an extent that new entrants to annotation jobs are often required to have undergraduate degrees. Many new graduates in technical disciplines such as computer science and engineering are recruited to entry-level jobs promising the allure of a career ladder in artificial intelligence.

Rather than pathways to gain technical skills and experience, the only opportunities for upward mobility are within the comically hierarchical management structures of the annotation industry: there's not much room at the top. Annotators who attempted to shift their career laterally into machine learning engineering or data science were told that they didn't have the right skills, expertise, and experience.

\subsection{Machines of flesh and blood: AI commoditizes knowledge work and separates it from its source}
Blackwell describes the behaviour of intelligent systems as the replay of (often subjective) human judgments, rather than the uncovering of a natural law~\cite{doi:10.1086/703871}. The latter `received view' of AI as extracting underlying principles from data comes to us via statistical modelling, which originated in helping 18th and 19th century astronomers make sense of their observations. These observations had minute errors due to the limitations of telescope technology, the human eye, and atmospheric conditions. Methods such as least squares regression allowed astronomers to account for such errors and `fit a line', as Carl Friederich Gauss did to estimate the location of the dwarf planet Ceres~\cite{lim_2021, doi:10.1177/002182867100200305}.

The ability to reconstruct `truth' from data led to wild extrapolations of statistical methods into other fields. The 19\textsuperscript{th} century Belgian astronomer Adolphe Quetelet, with his ``average man'', hypothesised that human individuals were imperfect renditions of God's will, that the perfect human was in fact the average of all humans (e.g., average height and weight), and that individuals who were closer to this average were, in some senses, more perfect~\cite{caponi2013quetelet,davis2013introduction,porter1986rise}. Quetelet's ideas were among the scientific rationales that fuelled the eugenics movement.

The foundations of statistical modeling as a way of taming the imperfections of nature still exert their objectivist influence on our thinking today. In believing that AI systems somehow capture a natural law that exists independently of the data annotators, we are hiding the exploitative labour that goes into AI systems. And by attributing undue agency to machines, we take it away from the people whose agency is captured and replayed through AI.

\subsubsection{Labour distancing is a very old trick}
The phrase ``labour distancing'' may call to mind Marx's theory that workers were separated – alienated – from their products by industrialisation, by machines that stripped them of their skills and reduced them to mere operators, and by the capitalist class that determined what to produce and how to produce it~\cite{marx1844comment}.

Yet the historical distancing of labour and product begins well before the industrial revolution. The European taste for sugar drove the dramatic rise of the transatlantic slave trade between the 16\textsuperscript{th} and 19\textsuperscript{th} centuries. The fact that most sugar was produced on remote plantations in the Caribbean that most Europeans could rarely see a painting of, let alone travel to, led to a convenient distancing of the sweet product from its bitter source. Thus, 18\textsuperscript{th} century abolitionists faced the challenge of bridging this mental distance. The pamphleteer William Fox evocatively depicts the West Indian sugar trade as a gruesome machine that transforms the bodies of slaves into sugar: \emph{``in every pound of sugar [...] we may be considered as consuming two ounces of human flesh''}~\cite{fox1791address}.

While formal chattel slavery, i.e., the legal ownership and trade of humans, was abolished in the Western world in the 19\textsuperscript{th} century, slavery did not disappear. Instead it reinvented itself and almost immediately reappeared in capitalist attire. The people of far away lands could not be directly owned. But by robbing their lands and privatizing their commons, they could be deprived of their ability to feed themselves, and thus forced to work on plantations. Or, ownership of their agricultural lands could simply be asserted through martial force, allowing them to continue working as before, but now as renters and taxpayers. They're not \emph{technically} slaves, so morally, you're in the clear. What a convenient trick!


European empires indulged themselves in this morality-laundering throughout Africa and Asia. The British, who still boast of their role in ``abolishing'' slavery, \emph{``found in the meek Hindu a ready substitution for the negro slave he had lost''} \cite{Emmer1986}, exporting over a million indentured labourers from India. The 1860 Dutch novel Max Havelaar, by Multatuli (pen name of Eduard Douwes Dekker), follows in the footsteps of Fox's pamphlets. Once called ``the book that killed colonialism''~\cite{pramoedya1999book}, it documents how the Dutch government's ``cultivation system'' required Indonesian farmers to grow sugar and coffee, rather than food, which caused extreme poverty and widespread starvation in Sumatra and Java.

The American empire followed a very similar strategy, creating de-facto colonies in South America in the 19\textsuperscript{th} and 20\textsuperscript{th} centuries, though the Americans insisted with bald hypocrisy that they did not have imperial aspirations. Corporations such as the United Fruit Company took pages straight out of the playbook of the East India Company, interfering with politics throughout central and South America, installing puppet dictators who privatized communal lands and sold them exclusive contracts. In Honduras, the use of toxic pesticides on banana plantations caused widespread disease and death, while authorities willingly turned a blind eye. In Guatemala, where United Fruit paid no taxes and controlled the government, as many as 200,000 people were ``disappeared'' during the civil war. In Colombia, banana workers and their families protesting peacefully for dignified working rights were brutally killed by machine gun fire in the 1928 Banana Massacre. These workers would be conveniently scapegoated by U.S. officials and United Fruit representatives as ``communists'', and murderous retaliation justified.

\emph{Coffeeland}, Augustine Sedgewick's magisterial history of Salvadoran coffee plantations, tells how the labour distancing of coffee became labour theft~\cite{sedgewick2021coffeeland}. As industrialists began realising that coffee can be used to improve the output of American factory workers, studies were conducted to determine with scientific precision how many coffee breaks, how often, and how long, were needed to optimise factory labour. Careful experimentation determined how a quantifiable serving of coffee could extend the workday by a quantifiable number of hours. Our modern practices of mid-morning and mid-afternoon coffee breaks stem from the findings of those studies.

The landmark 1956 Mitchell vs. Greinetz ruling established that coffee breaks were given to American workers to benefit employers, and workers must therefore be paid during these breaks~\cite{mitchellgreinetz1956, johnson1962workmen}. While the time and rights of North American coffee drinkers were protected, with benefits of increased productivity accruing to their employers, the time and rights of South American coffee plantation workers were steadily devalued. And there was tremendous incentive to do so: the cheap South American labour invested in growing coffee was amplified by ridiculous proportions in the productivity of expensive North American labour. In an argument mirroring Fox's equation of sugar to flesh, Sedgewick calculates that \emph{``The one and a half hours of work required to produce one pound of coffee in El Salvador became thirty hours of working time at [the American factory].''}

The illusion of automation and progress though labour distancing creates a powerful business advantage, even when the labour is only distanced by a few feet. America in the 20\textsuperscript{th} century witnessed the rise of the \emph{automat}, a vending machine restaurant~\cite{strauss_2021}. Customers saw a wall of coin-operated machines behind which lay hundreds of plates of freshly prepared food, such as meatloaf, mashed potatoes, and cherry pie. The success of this model, and its low price, was predicated upon hiding the team of cooks, dishwashers, technicians and waiters who operated the machines from the other side of the wall. At the end of the day, it was all a Wizard of Oz operation. Throughout history, when we peer into the depths of any seemingly autonomous machine, we see the faces of people peering back at us through the curtain.

\subsubsection{Artificial intelligence is distancing knowledge work}
Sugar and coffee are physical commodities, where units of human labour (e.g., people, time) can be mapped to units of produced output (e.g., weight), and the traces of their human source can be easily concealed. It is difficult to imagine how the same forces can shape labour distancing in knowledge work, but shape it they do.

Live music is arguably the first form of knowledge work to be distanced through modern information technology (visual art could always be physically removed from the producer, but musical performance could not). With the advent of recording technologies, such as vinyl phonograph records, it became possible to capture the labour of musicians, mass-produce it, and sell it cheaply, and for its consumption to be separated by time and distance from the artist. In the 1910s and '20s, demand for recordings by black musicians surged. Black musicians were inventing and reinventing the thrilling, modern sounds of blues and jazz, and white America wanted a slice of the action. Unsurprisingly, the history of this period is marred by the behaviour of the music industry, whose exploitative nature (for which it is notorious even today) was compounded by racism~\cite{lieb1981mother}. Underpaying artists, abusing them, refusing them credit, and stealing their compositions was standard practice.


August Wilson's \emph{Ma Rainey's Black Bottom}, a play about the influential early blues artist Gertrude ``Ma'' Rainey, depicts the cruel fate of black artists, whose hard-won success within white society was not enough to warrant fair treatment or even respect. If her crowds of white admirers could only experience Ma Rainey in physical person, confront her blackness and reconcile it against their own racism, she might have been afforded some of these dignities. But this uncomfortable hypocrisy could be avoided through the labour distancing afforded by recorded media. Wilson's Rainey says, \emph{``If you colored and can make them some money, then you all right with them. Otherwise, you just a dog in the street''}.

With AI powered by data annotators, the march of deskilling and labour distancing has arrived at mainstream knowledge work \cite{Pasquinelli2021}. Just as Caribbean sugar fuelled work in 18\textsuperscript{th} century Europe, and just as Salvadoran coffee fuelled factory work in 20\textsuperscript{th} century America, so too does the annotated data from the Global South fuel the knowledge work of the 21\textsuperscript{st} century West.

AI-driven tools promise to fix our grammar, improve our resumes, edit our photographs, design our presentations, drive our cars, curate our newsfeeds, and write our software code. \emph{Save your time}, sounds the siren song of artificial intelligence, \emph{be more productive}. The middlemen of AI software development therefore stand to profit handsomely from the same calculus of labour arbitrage that converted 90 minutes of Salvadoran plantation worker time to 30 hours of American factory worker time. 

Of course, this analogy has limits: even at its worst, the data annotation industry does not approach the monstrous coercion and suffering inflicted on colonial plantations. However, the parallels are obvious and it would be irresponsible to ignore them. Our moral goals have shifted over time. Fox agitated against slavery in the 18\textsuperscript{th} century, Multatuli against colonialism in the 19\textsuperscript{th}. In the 21\textsuperscript{st} century, we must agitate against labour distancing, starting with the misleading phrase `human-AI collaboration'.

Artificial Intelligence enables professional knowledge to be captured, replicated, mass-produced, commodified, and sold to consumers, stripped of the connection to the humans who provided the knowledge. To the long line of exploitative commodities: sugar, cotton, tobacco, cocoa, coffee, and mass media, we can now, terrifyingly, add knowledge itself.

\section{Synthetic and open datasets will not solve AI's labour exploitation problem}

There are ways of building AI systems without relying on large amounts of manually annotated data. First, we can generate synthetic datasets, as has been done very effectively with computer graphics (e.g.,~\cite{7785123}). For example, to build a system that detects someone's pose, such as sitting or standing, from a video feed, we can generate 3D renders of people in sitting and standing positions for training data. There's no need to gather this data by filming real people and then getting it manually annotated. Second, in scenarios such as playing digital games, an AI system can learn how to play simply by trying various actions within the game and receiving feedback about whether it is succeeding (e.g.,~\cite{schrittwieser2020mastering}). If an opponent is needed to play against, the system can play against itself. Third, if human data is needed, we can get it from the vast repositories of freely accessible data on the Internet: an AI can learn from the countless images, texts, videos, etc. already available.

These approaches are all promising, yet each has limitations which mean they cannot be applied in every situation. Synthetic data can only be generated in scenarios where we have clear underlying mathematical models of the domain. This is possible in computer graphics, where principles of vision and optics have been studied, distilled, and mathematized for centuries, and we have decades of experience in generating computer imagery of photorealistic scenes. Automatic feedback can only be generated in highly constrained settings such as games, where we have a straightforward signal of whether the system is doing well or not, e.g., whether the score is going up, or whether the system wins the game. Thus, there will always be situations where human-labelled data constitutes the best description of the desired behaviour of the system.


That leaves ``open data''. Many AI systems are trained on large datasets of openly-accessible text and images harvested from the Internet. This is merely grazing the digital commons, the argument goes, so there cannot be any harm.

The exploitation in open data is better concealed than in the annotation industry, but it is there. Earlier in this article, I wrote that ``the labour sharing and credit assignment of most software is fairly transparent'', the implication being that AI is an exception. Social media is another exception. Through psychological manipulation, these services transform Internet users into digital labourers, extracting work in the form of comments, posts, images, videos, etc ~\cite{fuchs2013digital}. With social media, as the saying goes, you're not the customer, you're the product. Beyond labour, the ecosystem of open data also enables control. As Zuboff notes in \emph{The Age of Surveillance Capitalism}, behavioural information such as posts, web searches, likes and dislikes are captured and used to anticipate and influence our future interests~\cite{zuboff2019age}.

Granted, not all training data comes from social media. Some online datasets are published with consent by the authors under permissive licenses (such as creative commons). However, because the size of the training dataset quite directly impacts the performance of the resulting model~\cite{https://doi.org/10.48550/arxiv.2203.15556}, there are powerful incentives against using only correctly-licensed data. Training datasets in practice draw from sources that span a spectrum of consent and exploitation.

There is also a legitimate argument against restricting training data to explicitly-licensed data: that such restriction would lead to bias. The resources, knowledge, and values around information sharing that might lead an individual or organisation to publish a dataset under a creative commons license are concentrated in the WWW: the Wealthy White World. And it would be a problem if our most advanced AI systems were trained on a dataset skewed towards the voices and interests of the wealthy white world. Thus in order to diversify the data and mitigate such biases, AI developers may need to draw upon resources published by underrepresented communities but which are not licensed for use in that manner. Underrepresentation, or paternalistic representation without consent: which is the greater evil? It is hard to say.

Is the answer, then, to advocate for more data from these communities gathered and shared in a responsible and consensual manner? It is not so simple: even the very concept of licensing promotes a certain Western worldview around knowledge, ownership, consent, and intellectual property rights. Just as land privatisation destroys conceptualisations of shared ownership such as the commons, and its nuanced cultural notions of the entitlement to and responsible use of natural resources, so too the Western notions of intellectual property rights have the propensity to destroy alternative (and perhaps more sustainable and equitable) conceptualisations of data~\cite{smith2021decolonizing}.

Exploitation goes beyond data. While this article focuses on the human labour of data annotation, for completeness it is necessary to briefly mention the human and environmental costs of the hardware that runs AI systems. There are many, but a few examples are the lithium for batteries, which comes from water-intensive salt mines in drought-stricken Bolivia; dysprosium and terbium for precision magnets from toxic and energy intensive mines in China; and tin from fish- and forest-killing Indonesian mines~\cite{crawford2018anatomy}. These material inputs for general-purpose computing devices, such as smartphones or laptops, must not be attributed solely to AI systems, as there is a great diversity of software these systems run, and certainly not all would qualify as AI. However, the attribution is easier in some cases, as with countertop voice assistants such as Amazon's Alexa, where the entire device is made to run a single AI system. And in the future, more and more of the software on our devices may directly contain or rely upon AI services to function.

\section{Artificial intelligence is a tool}
This article advocates for the metaphors AI is a \emph{tool}, or AI is an \emph{instrument}. Viewed in this way, the phrase ``human-AI collaboration'' becomes inconsistent with how we typically understand the term collaboration. We wouldn't say carpenter-hammer collaboration, for example, or surgeon-scalpel collaboration, or pianist-piano collaboration.

Is AI, as a tool, comparable to hammers, scalpels, or pianos? Language models such as GPT can write stories and poetry. Image generation models like DALL-E can generate beautiful images. What hammer could do that? In an effort to acknowledge the qualitative differences between the capabilities of AI and the capabilities of prior generations of tools, some coin new terms, such as ``supertool''~\cite{shneiderman2022human}. Others conceptualise AI as ``cognitive extenders'', which alter and extend our cognition and ways of information processing, placing it in the same category as inventions such as web search, maps, and writing~\cite{10.1145/3306618.3314238}.

But why do we feel it necessary to capture the difference between AI and a hammer (or for that matter, a computer)? It may be that our egos stand in the way of viewing AI as `just' a tool. Humans have a preoccupation with privileging intelligence that resembles their own, over other kinds of intelligence. The ethologist Frans De Waal has documented how the study of animal behaviour has been repeatedly set back by the insistence on human exceptionalism: how, when we test other animals for `intelligence', we test them in ways that are recognisable to us as intelligence but irrelevant to the circumstances of their evolution~\cite{de2016we}. But when we redesign tests in a less anthropocentric conceptualisation of intelligence, we are almost always surprised. Time and again, faculties considered to be unique to human intelligence — recognising faces, toolmaking, recollecting the past, planning the future, imagining alternative realities, etc. — have been demonstrated in other animals throughout the animal kingdom (i.e., not just in primates and corvids). Wasps can recognise each other's faces, for example.


If it is pride in AI's apparent resemblance to human intelligence that's blocking the tool metaphor, the solution is to question whether artificial intelligence truly resembles human intelligence. For AI to be a collaborator, we must ask: does it behave in all important respects as a human collaborator does? The answer, for now, is almost certainly `no'. Language AI, for example, is trained on language data only. Humans acquire language in a rich sensorimotor context: we learn by seeing, touching, feeling, hearing, and sensing the relationships between words and their meaning. But with no way to connect language with the real world concepts it refers to, language AI has no way to learn ``meaning'' and thus no ``understanding''~\cite{bender-koller-2020-climbing}. Nonetheless, this is likely to change in the future, with more advanced training techniques increasing the degree to which machine learning can be made embodied, continous, and multimodal.

But it is not just pride that drives the need to elevate AI to a higher status than other tools. There are also entrenched interests: academia, industry, and the media all gain from differentiating AI from other tools. Academics need to attract grant funding, citations, and press attention. Industry needs to capture investment, thought leadership, customers, and news cycles. The media need to attract eyeballs.

Of the interested parties, it is the media that needs to be held most accountable for shaping the perceptions of lay observers. Not only does media ``shape, mediate, and amplify expectations surrounding artificial intelligence'', but also perpetuates a mythology of AI as being widely disruptive across society and applicable to nearly any problem~\cite{doi:10.1177/1464884920947535}. Moreover, one of the principal ways in which media presents progress in AI is as a ``competition'' that pits humans against machines.

Often, the media co-opts prestigious public intellectuals to legitimise the message, with little regard to the precise domain of their expertise. Galanos gives such people the euphemistic term ``expanding experts'', in the sense that they expand their commentary to include topics beyond their expertise, such as AI~\cite{doi:10.1080/09537325.2018.1518521}. Experts who reach beyond their domain are a problem, because their public position influences the opinions of lay audiences, academics from other domains, and policy documents. This is not unique to artificial intelligence discourse. Bender demands bluntly~\cite{bender_2022}: \emph{``every puff piece that fawns over its purported ``intelligence'' lends credence to other applications of ``AI'' [...] We should demand instead journalism that refuses to be dazzled by claims of ``artificial intelligence'' [...] It behooves us all to remember that computers are simply tools.''}

The irony is that despite the intelligence of AI being largely derived from the performance of human intelligence, its ability to replay it is nothing like human intelligence. This is not to say that AI is better or worse, as those comparators imply a reductionist continuum of intelligence, a single scale along which all intelligences can be placed. It is merely different. But it is different in ways that ought to rule out collaborator, partner, and teammate as appropriate metaphors.

\section{Beyond the collaboration metaphor}

It is worth revisiting the source of the human-AI collaboration metaphor. Where does it come from? What is it comparing against? If there are so many problems with it, why does the community use it in the first place?

The metaphor of AI as collaborator comes from a genuine desire to support and empower users of these systems. It has been developed as a response to the narrative of apocalyptic automation, where the human has been cut out of the loop, replaced, reduced, dehumanised, obviated. Instead of a future where AI takes our jobs, and takes control from us, the human-AI collaboration metaphor imagines a future where we work together with AI systems, each building on the unique strengths of the other. This is a sentiment that has the best of intentions, and I am not criticising those who use the metaphor in this way.

But we can do better. We can empower users without disenfranchising the nameless labourers of AI in the process.

The title of this article takes inspiration from Fiona Ssozi-Mugarura and colleagues at the University of Cape Town, whose paper \emph{Enough with `In-The-Wild'} critiques the use of the phrase `in-the-wild' to describe real-world (as opposed to laboratory) studies, which comes to us from an era of anthropological research where societies in Africa were considered `wild'~\cite{10.1145/2998581.2998601}. By titling this article after theirs, I am suggesting that we must similarly critique phrases such as ``human-AI collaboration'', or ``human-AI partnership''.


Terminology matters. It matters particularly in AI research, where words such as ``decide'', ``goal'', ``plan'', ``understand'', and ``collaborate'' are picked out of ordinary use and repurposed for specialised technical use in an uncritical manner~\cite{agre1997toward}. Shifting terminology to make our values explicit has much precedent in the field of human-computer interaction. Indeed the field itself was consciously renamed from ``computer-human interaction'' to ``human-computer interaction'', to put the human first.\footnote{The vestiges of the old name can still be seen in our most prestigious conference, \emph{CHI}.} It turns out there are humans on both sides of the computer. We must put them \emph{all} first.


We're not going to solve all the problems discussed in this article through a change in terminology. But the HCI research community can at least stop some of the harms of labour distancing by acknowledging that human-AI collaboration is really human-human collaboration, distanced and disguised, in a way that is qualitatively distinct from other forms of human-computer interaction and software use. When we choose metaphors around which to structure our theories of AI design, we must begin from the equitable position that AI is a tool, not a collaborator or a partner. And if we must invoke the idea of a collaborator or partner, it must be rooted in acknowledgement of the skills and knowledge captured and replayed by the AI system, and the people from which it comes.

\epigraph{I am not your data, nor am I your vote bank,\\
I am not your project, or any exotic museum object,\\
I am not the soul waiting to be harvested}{Abhay Flavian Xaxa, \emph{I Am Not Your Data}}

\bibliographystyle{ACM-Reference-Format}
\bibliography{sample-base,references}










\end{document}